\documentclass[useAMS,usenatbib]{mn2e}

\usepackage{latexsym,graphicx,natbib}
\usepackage{amsmath}

\newcommand\simless{{\thinspace \rlap{\raise 0.5ex\hbox{$\scriptstyle  {<}$}}
    {\lower 0.3ex\hbox{$\scriptstyle  {\sim}$}} \thinspace }}  
\newcommand\simgreat{{\thinspace \rlap{\raise 0.5ex\hbox{$\scriptstyle  {>}$}}
    {\lower 0.3ex\hbox{$\scriptstyle  {\sim}$}} \thinspace }}  

\newcommand\msun{\, \rm M_\odot} 
 
\newcommand\mpc{{\, \rm mpc}}
\newcommand\pc{{\, \rm pc}}
\newcommand\kpc{{\, \rm kpc}}
\newcommand\kms{{\, \rm km\,s^{-1}}}
\newcommand\yr{{\, \rm yr}}

\newcommand\mbh{M_{\rm MBH}}
\newcommand\mim{M_{\rm IBH}}
\newcommand\mstar{m_{\rm *}}
\newcommand\mtot{M_{\rm MBH} + M_{\rm IBH}}
\newcommand\tpr{T_\varpi}
\newcommand\tgw{T_{\rm GW}}
\newcommand\mas{\,\rm mas}
\newcommand\einn{e_{\rm inn}}
\newcommand\eout{e_{\rm out}}
\newcommand\Pinn{P_{\rm inn}}
\newcommand\Pout{P_{\rm out}}

\def\aj{AJ}%
\def\apj{ApJ}%
\def\apjl{ApJL}%
\def\aap{A\&A}%
\def\mnras{MNRAS}%
\def\nat{Nature}%
\def\prd{Phys.Rev.D.}

\title[The Galactic Centre star S2 as dynamical probe]{The Galactic
  Centre star S2 as a dynamical probe for intermediate-mass black  holes}

\author[A. Gualandris, S. Gillessen and
   D. Merritt]{A. Gualandris$^{1}$\thanks{E-mail:
     alessiag@mpa-garching.mpg.de}, S. Gillessen$^{2}$ and
   D. Merritt$^{3}$\footnotemark[1]\\ $^{1}$Max-Planck Institut
   f\"{u}r Astrophysik, Karl-Schwarzschild-Str. 1, D-85748 Garching, Germany
   \\ $^{2}$Max-Planck Institut f\"{u}r Extraterrestrische Physik,
   Giessenbach-Str., D-85748, Garching, Germany\\ $^{3}$Department of
   Physics and Center for Computational Relativity and Gravitation,
   Rochester Institute of Technology, \\ 54 Lomb Memorial Drive,
   Rochester, NY}

\begin{document}

\date{}

\pagerange{\pageref{firstpage}--\pageref{lastpage}} \pubyear{2010}

\maketitle

\label{firstpage}

\begin{abstract}
We study the short-term effects of an intermediate mass black hole
(IBH) on the orbit of star S2 (S02), the shortest-period star known to
orbit the supermassive black hole (MBH) in the centre of the Milky
Way. Near-infrared imaging and spectroscopic observations allow an
accurate determination of the orbit of the star. Given S2's short
orbital period and large eccentricity, general relativity (GR) needs
to be taken into account, and its effects are potentially measurable
with current technology.  We show that perturbations due to an IBH in
orbit around the MBH can produce a shift in the apoapsis of S2 that is
as large or even larger than the GR shift. An IBH will also induce
changes in the plane of S2's orbit at a level as large as one
degree per period. We apply observational orbital fitting techniques to
simulations of the S-cluster in the presence of an IBH and find that
an IBH more massive than about $1000\msun$ at the distance of the
S-stars will be detectable at the next periapse passage of S2, which
will occur in 2018.

\end{abstract}


\section{Introduction}
The innermost arcsecond of the Milky Way harbours a cluster of young
massive stars (the S-star cluster) in eccentric orbits around the
supermassive black hole (MBH). Near-infrared observations of the
cluster allow a precise determination of the trajectories of about 20
stars. These can be used to derive fundamental parameters like the
mass of the MBH and the distance to the Galactic centre, as well as to
constrain the gravitational potential and test predictions from
general relativity (GR) \citep{rubilar2001,zucker2006}.  Since the
collapse of a giant molecular cloud at the distance of the S-cluster
is prevented by the tidal force of the MBH, it is believed that either
the S-stars formed in-situ via a non-standard process or that they
originated outside the black hole's sphere of influence and then
migrated to their current location.  Among the in-situ models are the
two disks model \citep{lock09} and the eccentric instability model
\citep{madigan09}.  The two main models suggested for the transport of
the stars are the cluster infall scenario \citep{gerhard01}, also
aided by an intermediate-mass black hole (IBH) \citep{hm03, kim04,
  levin05, mgm09, fujii09, fujii10}, and the binary disruption
scenario \citep{gq03,perets07,perets09}. The properties and efficiency
of these models are discussed in \citet{PG10}.

The star with the shortest orbital period ($\sim 15\yr$), called
S2 \citep{schodel2002, ghez03}, has a semi-major axis $a =
(0.1246\pm 0.0019)''$ and an eccentricity $e = 0.8831 \pm
0.0034$ \citep{gillS2}.  Adopting a distance to the Galactic
centre $R_0 = 8.28\kpc$, $a \sim 5.0018\mpc$, while the
periapsis and apoapsis distances are, respectively, $r_p =
0.585\mpc$, $r_a = 9.419\mpc$ .  For this set of orbital
elements, GR precession is measurable with current
instrumentation on a time-scale of about 10 years
\citep{Gillessen2009}.

The relativistic (prograde) advance of the periapse angle is given in
the case of a non-rotating black hole by
\begin{equation}
\Delta\varpi = \frac{3\pi}{1-e^2} \frac{R_S}{a} = \frac{6\pi G}{c^2} \frac{\mbh}{a(1-e^2)}
\end{equation}
per radial period  \citep{weinberg72},
where $R_S = 2G\mbh/c^2$ is the Schwarzschild radius of the black
hole.  For the Galactic centre black hole, $R_S \approx
4.1\times10^{-7}\pc$ for an assumed mass of $\mbh=4.3\times10^6\msun$
\citep{gillS2}.
As a result of the orbit's precession, there is a displacement in
the star's apoapse position that is given by
\begin{equation}
\Delta r_a \approx a (1+e) \Delta\varpi \approx 
\frac{6\pi G\mbh}{c^2(1-e)}
\end{equation}
again per radial period; note that this expression is independent of the
semi-major axis.
As seen from Earth, this shift corresponds to an angle on the sky of
\begin{equation}
\label{eq:wein}
\Delta \Theta_a \sim 0.097{\mas} \left(\frac{1}{1-e}\right)
\left(\frac{\mbh}{4.3\times10^6\msun}\right) \left(\frac{8.28\kpc}{R_0}\right)
\end{equation}
which amounts to $\sim 0.83\mas$ for star S2.

A complicating factor is the likely presence of a distributed mass
within S2's orbit, consisting of stars and stellar remnants.  Assuming
that the mass density follows $r^{-\gamma}$, with $r$ the distance
from the MBH, the advance, per radial period, of orbital periapse due
to the distributed mass is
\begin{equation}
\Delta\varpi \approx  2\pi \frac{M_\star(a)}{\mbh}\sqrt{1-e^2}F(\gamma)
\end{equation}
in the retrograde sense;
here $M_\star(r)$ is the distributed mass enclosed within radius
$r$ and $F=(3/2,1)$ for $\gamma=(0,1)$ \citep[e.g.][]{MAMW2010}.
Setting $F\approx 1$, this implies, for S2, a shift on the sky of
\begin{equation}
\Delta\Theta_a \approx 0.69 \mathrm{mas} \left(\frac{10^3 M_\star(a)}{\mbh}\right).
\end{equation}
This is comparable with the relativistic precession if the distributed
mass within S2's orbit is $\simgreat 10^{-3}\mbh$.
Disentangling these two sources of precession will be difficult, requiring
measurements of the radial velocity of S2 near periapse passage
\citep[e.g.][]{saha10}.

Non-spherically-symmetric perturbations, if present, can affect not
just the periapse angle but also the angular momentum of S2's orbit,
resulting in changes in $e$ and in the orientation of its orbital
plane.  Potentially the largest source of such perturbations is a
second massive black hole orbiting somewhere near S2.  In fact, the
orbits of the S-stars are consistent with the long-term presence of an
IBH in their midst, with mass $\sim 10^3\msun$ \citep{mgm09,GM2009}.
While an IBH would also contribute to the periapse advance of S2, its
key signature would be a change in S2's orbital angular momentum.

In this work, we combine high-accuracy $N$-body simulations 
(\S2) with orbital fitting techniques (\S3) to investigate the
observable effects of an IBH on the orbit of star S2 over a time-scale
of a few orbital revolutions.
In \S4 we discuss the other types of perturbation that could 
potentially induce 
changes in S2's orbital angular momentum and compare them with
perturbations from an IBH.
\S5 sums up.

\section{Initial models and numerical methods}
We consider $N$-body models that include a MBH, an IBH and the S-star
cluster.  The initial conditions for the stars are derived from the
orbital elements given by \citet{Gillessen2009}. For star S2, we take
the improved elements from \citet{gillS2}.  Of the 28 stars for which
those authors provide orbital elements, we exclude the six stars (S66,
S67, S83, S87, S96, S97) which likely belong to the clockwise disk
\citep{genzel03, paumard06} and star S111 which appears to be unbound.
We are left with a sample of 21 stars with well defined orbits, for
which we determined positions and velocities at 2008 AD from the
classical elements.  The masses of the S-stars were set to $10\msun$
\citep[e.g.][]{eisen05} except for star S2 for which a value of
$20\msun$ was adopted \citep{martins08}.

The IBH is placed on a Keplerian orbit around the MBH.  We adopt four
different values for the mass ratio of the black hole binary $q \equiv
\mbh/\mim = (1.0\times10^{-4}, 2.5\times10^{-4}, 5.0\times10^{-4},
1.0\times10^{-3})$, five values for the semi-major axis $a = (0.3, 1,
3, 10, 30)\mpc$, four values for the eccentricity $e = (0, 0.5, 0.7,
0.9)$ and twelve choices for the direction of the orbit's angular
momentum vector (the same set as in \citet{GM2009}), for a total of
960 sets of initial conditions.  The IBH begins from orbital periapsis
in all cases.  The initial value of the mean anomaly is likely to be
unimportant in all cases for which the orbital period of the binary is
much smaller than the integration time ($50\yr$), i.e. $a=0.3, 1,
3\mpc$. For $a=30\mpc$, only part of the IBH orbit is sampled by the
integration, and in principle the choice of the initial position might
have an effect on the interaction with S2. However, at a distance of
$30\mpc$ the IBH is completely outside S2's orbit for $e \simless
0.7$, and there is essentially no detectable signature, as shown
below. The only set of simulations for which the initial mean anomaly
of the IBH might be important is the $a=10\mpc$, which corresponds to
an orbital period of about $43\yr$ for the IBH. We perform an extra
set of simulations for $a=10\mpc$ starting the IBH at apoapsis rather
than periapsis and we compare the results in the two cases.

We advanced each $N$-body system in time using the AR-CHAIN code
\citep{mm08}, a recent implementation of the algorithmic
regularisation method that is able to reproduce the motion of tight
binaries for long periods of time with extremely high precision.  The
code combines the use of the chain structure, introduced originally by
\cite{ma93}, with a new time transformation to avoid singularities and
achieve high precision for arbitrary mass ratios.  Note that we
self-consistently follow not just the interactions of S2 with the MBH
and IBH, but all other interactions as well, including star-star
interactions.  The integration interval was 50 years, during which
time S2 performs three full orbits.

\begin{figure}
  \begin{center}
    \includegraphics[width=8cm]{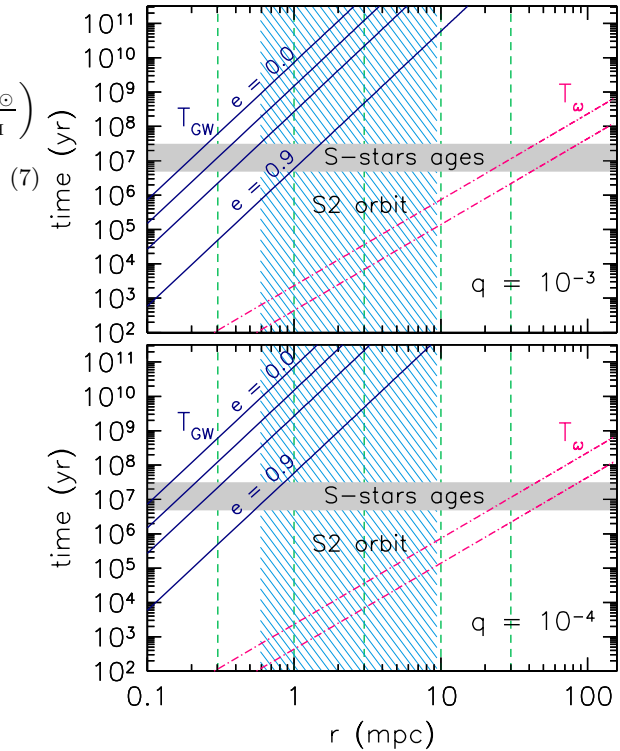}
  \end{center}
  \caption{Time-scales associated with orbital evolution in our
    models.  Solid lines show the GW time-scale, Eq.~(\ref{eq:tgw}),
    for a black hole binary with $q = 10^{-4}$ (top) and $q = 10^{-3}$
    (bottom), for four different values of the eccentricity $e=0, 0.5,
    0.7, 0.9$. Dashed-dotted lines show the GR precession time-scale
    for two different values of the eccentricity: $e = 0$ (upper line)
    and $e = 0.9$ (lower line). The vertical dotted lines represent
    the adopted values for the binary initial semi-major axis. The
    filled grey region indicates the estimated ages of the S-stars
    while the striped area shows the radial range of S2's orbit.}
  \label{fig:times}
\end{figure}

The AR-CHAIN code includes relativistic corrections to the
accelerations up to 2.5 post-Newtonian order for all interactions
involving the MBH particle.  General relativistic advance of the
periapse, which operates on a time-scale
\begin{eqnarray}
  \tpr & \equiv &\left|\frac{\Delta\varpi}{2\pi P}\right|^{-1} =
 \frac{2 \pi c^2 \left(1-e^2\right) a^{5/2}}{3 \left(G \mbh \right)^{3/2}} \nonumber \\ 
       & \approx &  1.3\times10^5 {\rm yr} \left(\frac{a}{5\mpc}\right)^{5/2} \left(\frac{4.3\times10^6\msun}{\mbh}\right)^{3/2}
\left(1-e^2\right)
\end{eqnarray}
is accounted for by the 1PN and 2PN terms.  The dissipative term
arising from the emission of gravitational waves is accounted for by
the 2.5PN term; this term is potentially important for the IBH, for
which the associated coalescence time-scale is
\begin{eqnarray}
\label{eq:tgw}
  \tgw & = & \frac{5}{256 F(e)} \frac{c^5}{G^3} \frac{a^4}{\mu
    \left(\mtot\right)^2} \nonumber\\ & \approx &
  \frac{1.96 \times 10^{13} \yr}{F(e)} \left(\frac{a}{5\mpc}\right)^4
  \left(\frac{4.3\times10^6\msun}{\mbh}\right)
  \left(\frac{10^3\msun}{\mim}\right)\nonumber\\ & & \times
  \left(\frac{4.3\times10^6\msun}{\mtot}\right)
\end{eqnarray}
where
\begin{equation}
  F(e) = \left(1-e^2\right)^{-7/2} \left(1 + \frac{73}{24}e^2 + \frac{37}{96} e^4\right) \nonumber
\end{equation}
and
\begin{equation}
  \mu = \frac{\mbh\mim}{\mtot} \approx \mim \nonumber
\end{equation}
is the reduced mass of the IBH/MBH binary.  The orbital decay
time-scale for the black hole binary is shown in Figure~\ref{fig:times}
for the two extreme values of the mass ratio and the four adopted
values of the initial eccentricity.  This time-scale is always much
longer than our integration interval of $\sim 50$ yr.  In addition, it
is longer than the main-sequence lifetime ($\sim 10^7$ yr) of a $20$
solar mass star for all initial configurations excepting the cases $a
= 0.3\mpc$ and $e \geq 0.7$.  In the former runs, it is justified to
associate our initial parameters for the IBH/MBH binary with the
parameters at some much earlier time, e.g. the epoch preceding
formation of the S-stars.  In the latter runs, the orbit of the IBH at
some much earlier time would have been larger and/or more eccentric.
The maximum relative variation of the binary semi-major axis in the $N$-body
integrations is $\Delta a / a \sim 10^{-3}$ while the absolute variation
of the eccentricity is $\Delta e \sim 10^{-2}$.

\section{IBH perturbations}

In the Schwarzschild metric, the argument of periapse $\varpi$ evolves
due to in-plane precession.  The two remaining angles that define the
orientation of the orbit, $i$, the inclination, and $\Omega$, the
position angle of the ascending node, are fully conserved in the
relativistic two-body problem.\footnote{We follow the standard
  practise of using the plane of the sky as the reference plane for
  defining $(\Omega,i)$.}  The semi-major axis and eccentricity are
conserved at the 1PN level, and we expect very small deviations due to
higher order PN corrections in the integrations \citep{soffel89}.  In
the limit of small star-to-black hole mass ratio, the semi-major axis
and eccentricity in the PN approximation are given by \citep{soffel89}
\begin{equation}
a = \frac{-GM}{2\mathcal{E}} \left[ 1 +
  \frac{7}{2} \frac{\mathcal{E}}{c^2}\right]
\end{equation}
\begin{equation}
e = \sqrt{1 + \frac{2\mathcal{E}}{G^2 M^2}
  \left(1+ \frac{17}{2} \frac{\mathcal{E}}{c^2}\right)
  \left(\mathcal{J}^2 + 2 \frac{G^2M^2}{c^2}\right) }
\end{equation}
where
\begin{equation}
\mathcal{E} = \frac{1}{2}v^2 - \frac{GM}{r} + \frac{3}{8}
\frac{v^4}{c^2} + \frac{GM}{2rc^2} \left[3v^2 + \frac{GM}{r}\right]
\end{equation}
is the specific post-Newtonian energy and
\begin{equation}
\mathcal{J} = |\overrightarrow{x} \times \overrightarrow{v}|
\left[ 1 + \frac{1}{2} \frac{v^2}{c^2} + \frac{3GM}{rc^2} \right]
\end{equation}
the specific angular momentum.  Here, $\overrightarrow{x}$ and
$\overrightarrow{v}$ are the relative position and velocity vectors
between the star and the MBH, $M$ is the total mass, and $c$ is the speed
of light.

The presence of the other S-stars introduces small deviations from
spherical symmetry in the gravitational potential but the effect on
the orbital elements over these short time-scales is negligible,
as we show in Section \ref{sec:torque}.
Therefore, the variations that we observe in the orbital elements of star S2
in our simulations,
namely the semi-major axis, eccentricity, inclination, and position
angle of the ascending node, can be attributed to perturbations from
the IBH.

\begin{figure}
  \begin{center}
    \includegraphics[width=4cm]{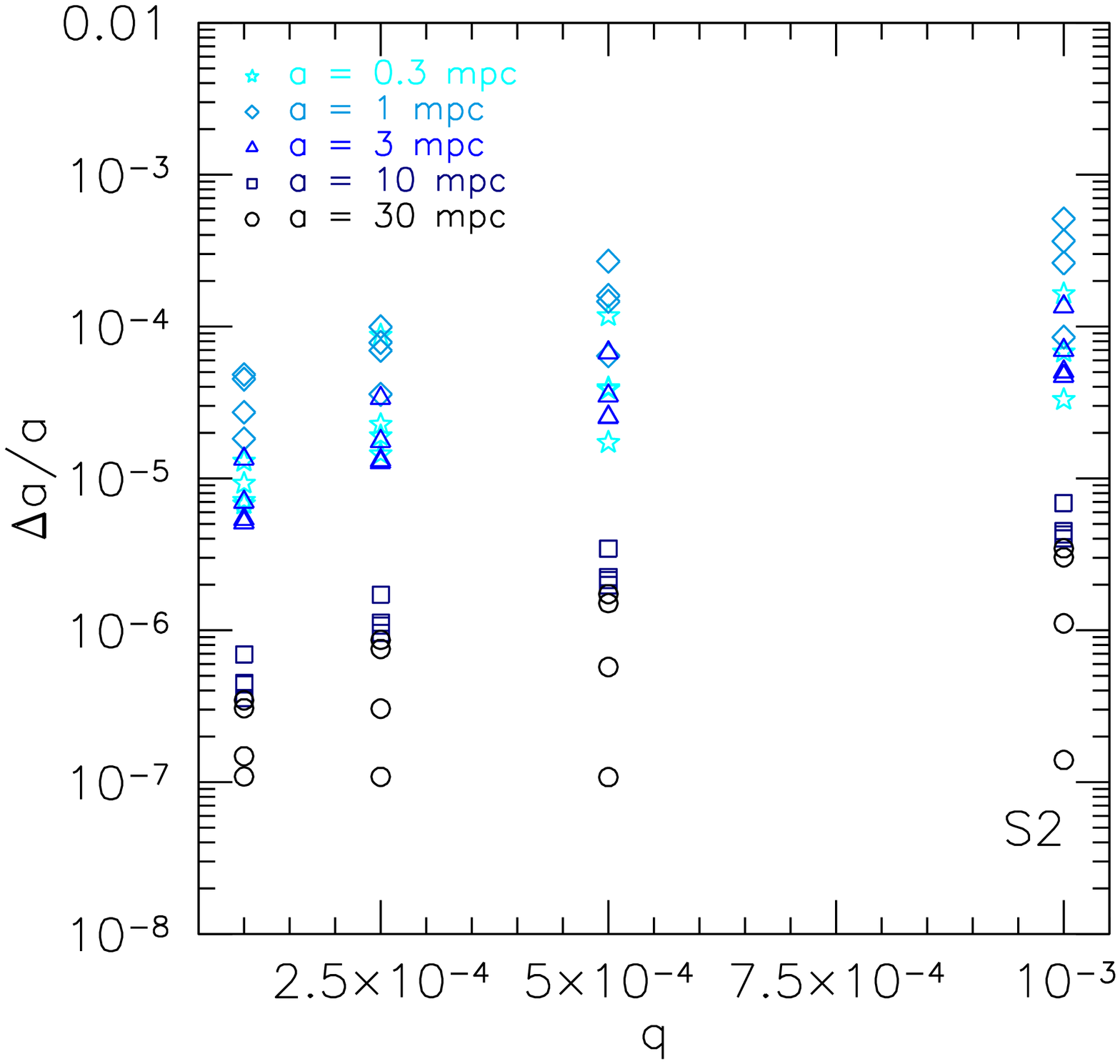}
    \includegraphics[width=4cm]{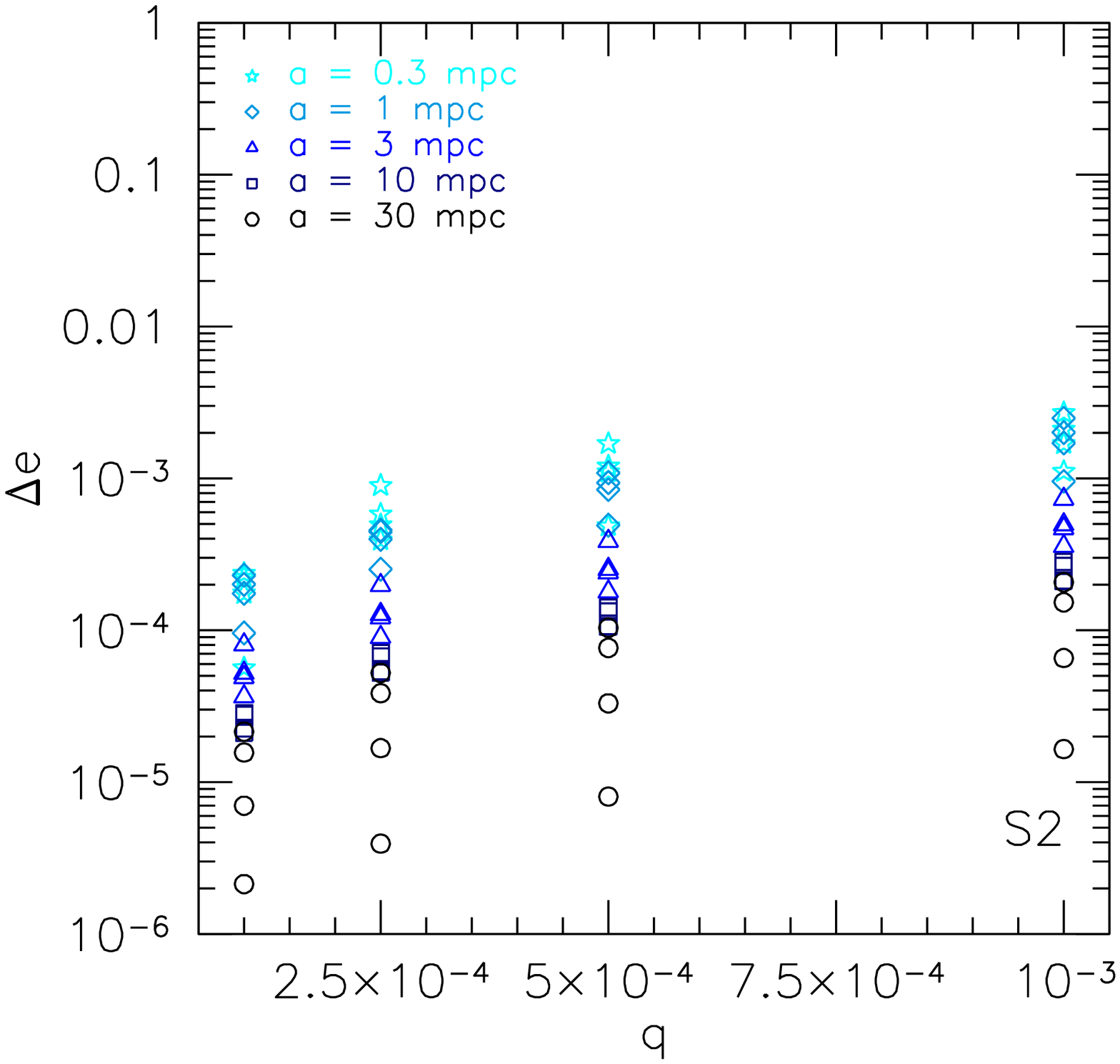}
    \includegraphics[width=4cm]{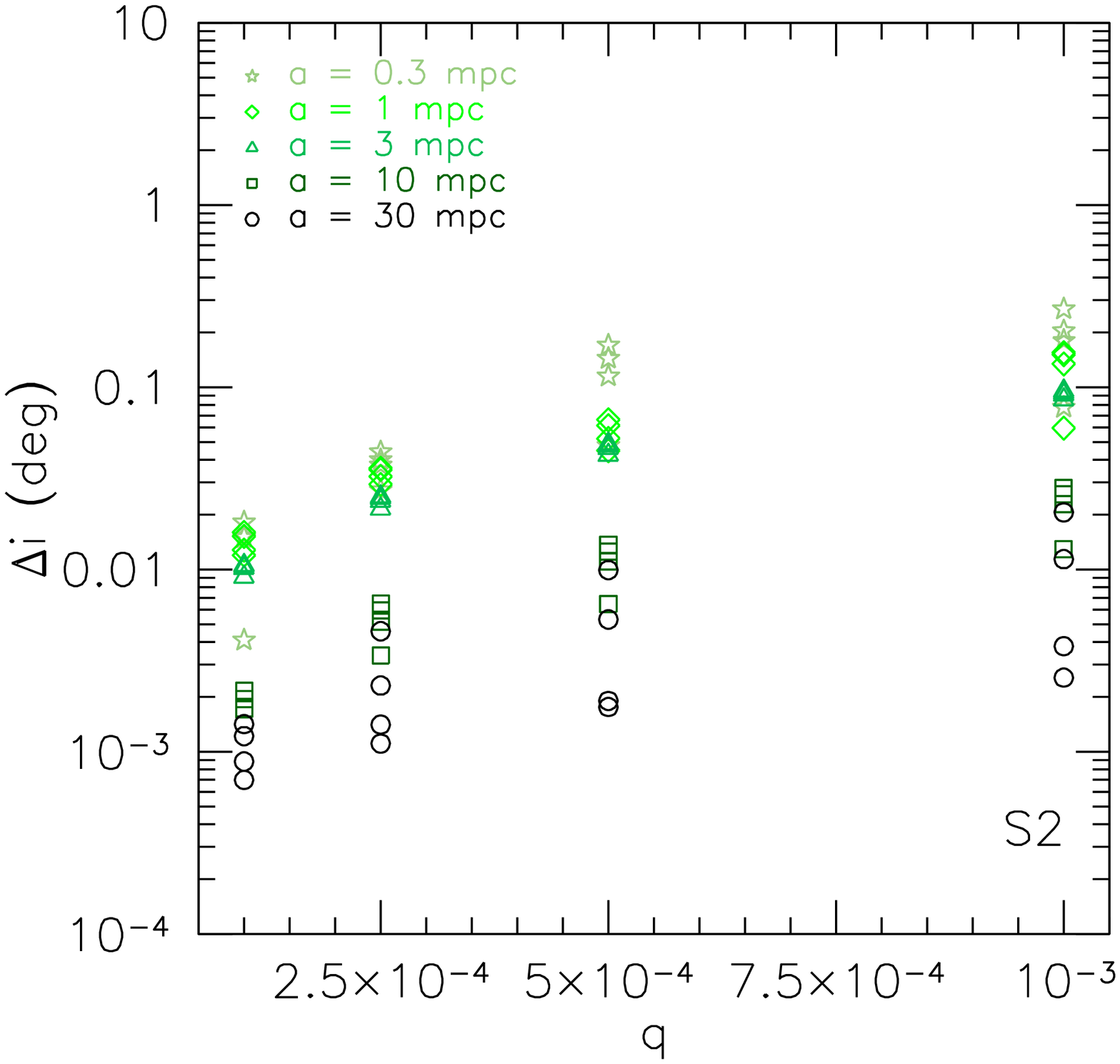}
    \includegraphics[width=4cm]{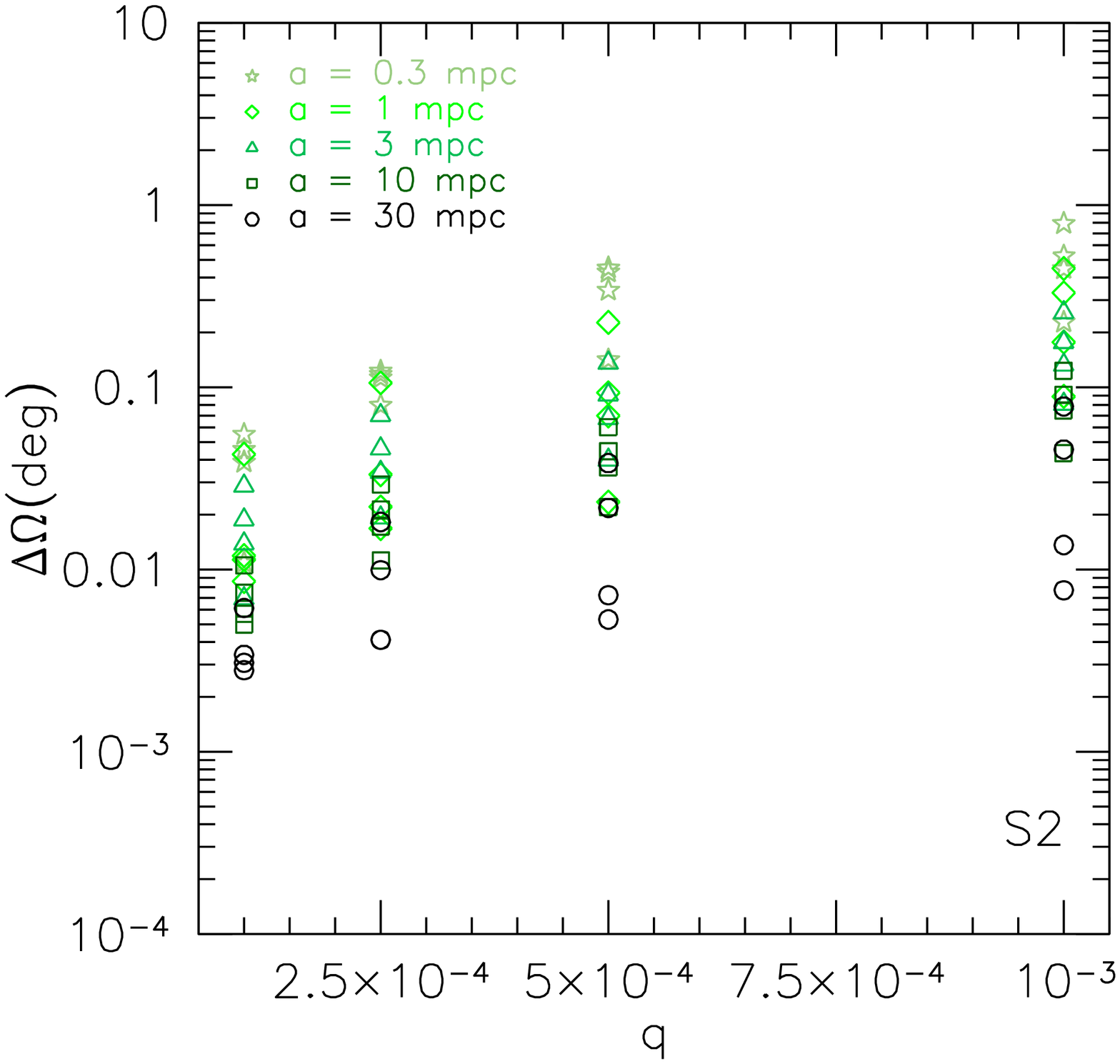}
    \includegraphics[width=4cm]{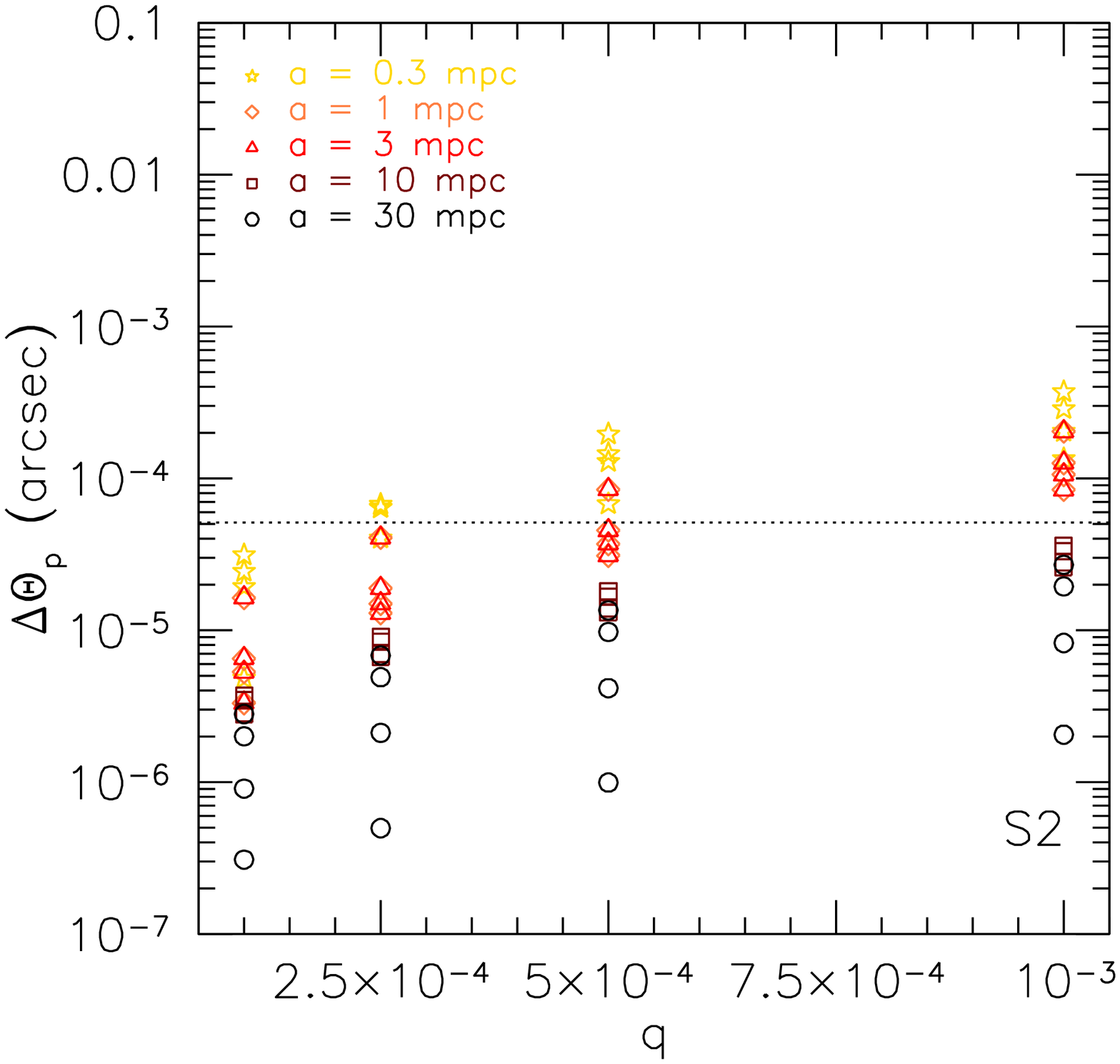}
    \includegraphics[width=4cm]{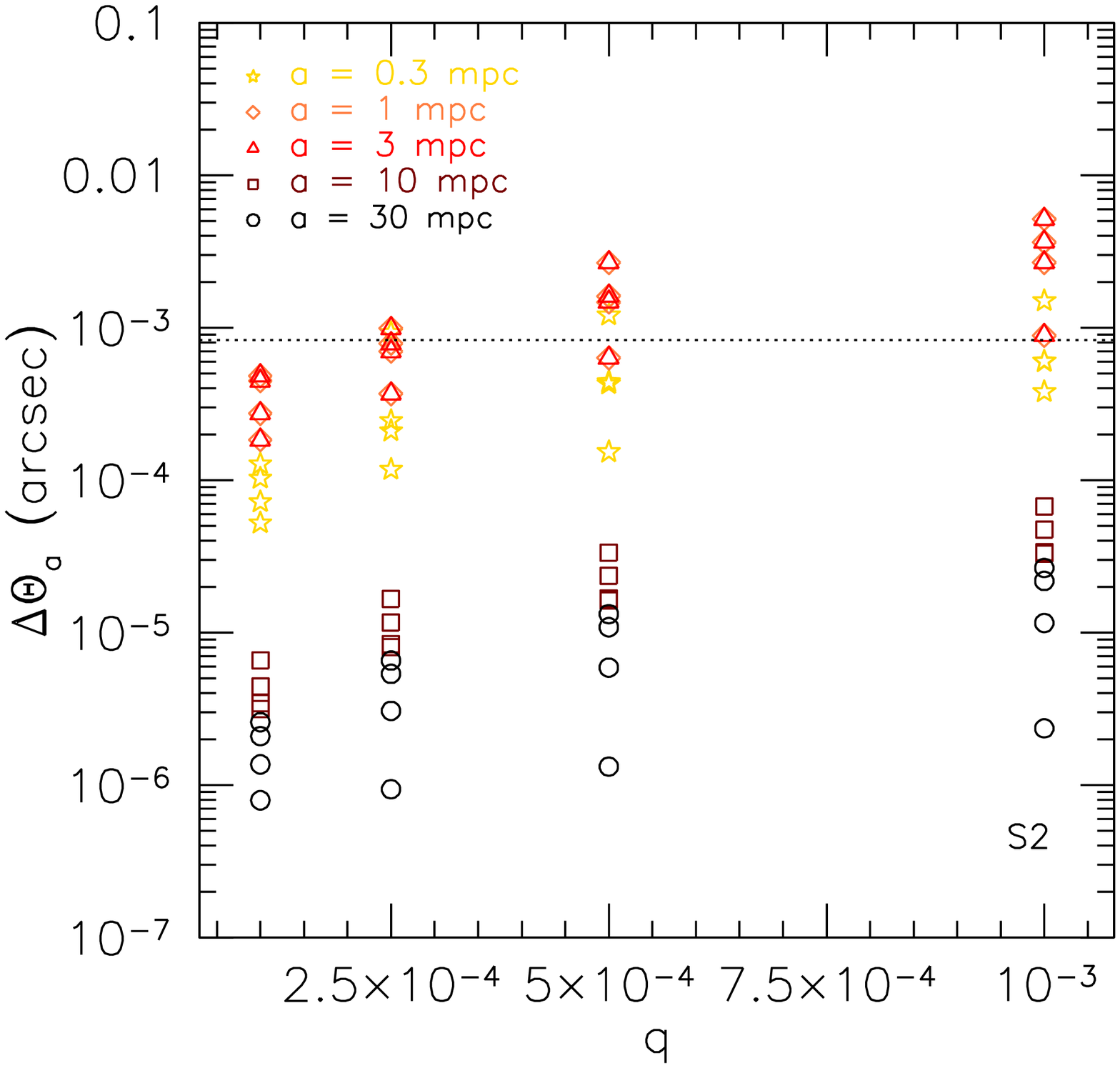}
  \end{center}
  \caption{Average changes in the orbital elements (semi-major axis,
    eccentricity, inclination, position angle of the ascending node,
    periapsis and apoapsis) of star S2 over one full orbit, versus the
    mass ratio of the black hole binary.  Different symbols are for
    different initial semi-major axes of the binary. Each point is an
    average over the 12 orientations of the IBH/MBH orbital angular
    momentum vector. The dotted lines represent the GR shift in the
    periapse and apoapse.}
  \label{fig:change}
\end{figure}

Figure~\ref{fig:change} summarises the changes in the orbital elements
of S2 found in the $N$-body integrations.  Plotted are the variations
over one revolution averaged over the twelve different orientations of
the initial IBH/MBH orbit.
The dotted lines in the periapsis and apoapsis panels indicate the
variations due to GR. The shift in the periapse corresponds to an
observable angle $\Delta \Theta_p = \Delta \Theta_a
\left(1-e\right)/\left(1+e\right)$, where $\Delta \Theta_a$ is defined
in Eq.~\ref{eq:wein}.  Note that the variations in the inclination $i$
and position angle $\Omega$ of the ascending node reach values close
to 1 degree for the most massive IBHs considered. This is of the same
order as the current observational accuracy ($\sim 0.7\,\rm deg$). In
50 years, this value will drop to about $0.4\,\rm deg$, assuming there
are no technological improvements.

While precession induced by the PN terms is restricted to the orbital
plane, an IBH induces more general changes in the orbital elements,
including changes in the direction of the orbital angular momentum
vector. We measure the latter via the angle
\begin{equation}
\cos \phi = \left( \frac{\overrightarrow{L_i} \cdot \overrightarrow{L_f}}{L_i\,L_f}\right)\,.
\end{equation}
\begin{figure}
  \begin{center}
    \includegraphics[width=8cm]{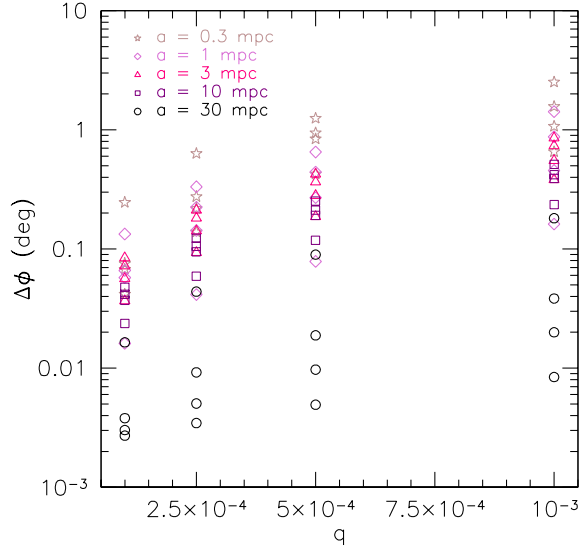}
  \end{center}
  \caption{Mean variation of the orbital plane of star S2 as a
    function of the binary mass ratio. For each combination of binary
    mass ratio, semi-major axis and eccentricity, the results are
    averaged over the 12 orbital orientations of the IBH.}
  \label{fig:phi}
\end{figure}
Figure~\ref{fig:phi} plots $\phi$ for all the runs, after averaging
over the 12 different IBH orientations.

The changes in the orbital plane of S2 are larger for more massive
IBHs and can reach values of $\sim 1$ degree for $q =
10^{-3}$. Out-of-plane motion is also affected by the size of the
MBH/IBH binary orbit such that changes are largest for $a \simless
10\mpc$.

The three-body system composed of MBH, IBH and S2 is reminiscent of a
Kozai triple \citep{kozai62}.  However, the changes that we observe in
the orbital elements are not generally attributable to the Kozai
mechanism.  Kozai oscillations can be induced if (i) the MBH - S2 -
IBH system can be regarded as a hierarchical triple, with
well-separated orbital periods for the inner and outer orbit; (ii) the
period predicted for the Kozai oscillations is shorter than that of
any other precessional period, in this case, GR precession; (iii) the
outer orbit is largely inclined with respect to the inner orbit.  The
first condition is only satisfied in the runs with $a = 10, 30\mpc$.

The timescale for Kozai oscillations can be written as \citep{KH07}
\begin{equation}
\label{eq:tkozai}
T_K = \frac{4 \mathcal{K}}{3\sqrt{6}\pi} \frac{\Pout^2}{\Pinn}
\frac{\mbh + \mstar}{\mim} \left(1-\eout^2\right)^{3/2}
\end{equation}
where $\Pinn$ and $\Pout$ are the period of the inner and outer
binary, $\eout$ is the eccentricity of the outer orbit and
$\mathcal{K}$ is a numerical coefficient which depends only on the
initial values of the relative inclination angle $\alpha$, the inner
binary eccentricity $\einn$ and the inner binary argument of periapsis
$\varpi$.  The maximum eccentricity $e_{\rm max}$ attained by the
inner binary is also a function of the initial values of $\alpha$,
$\einn$ and $\varpi$ only.  
\begin{figure}
  \begin{center}
    \includegraphics[width=4cm]{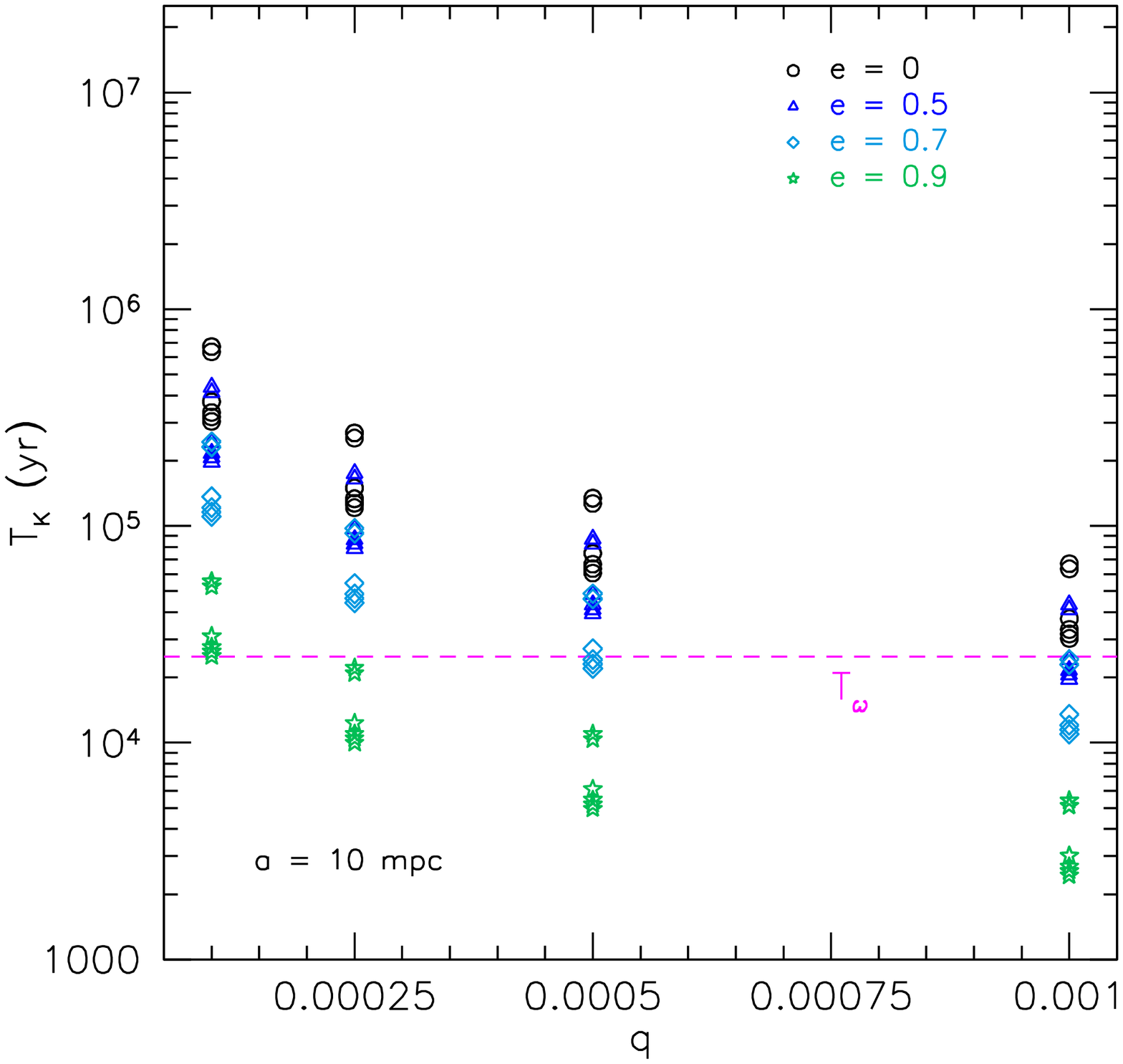}
    \includegraphics[width=4cm]{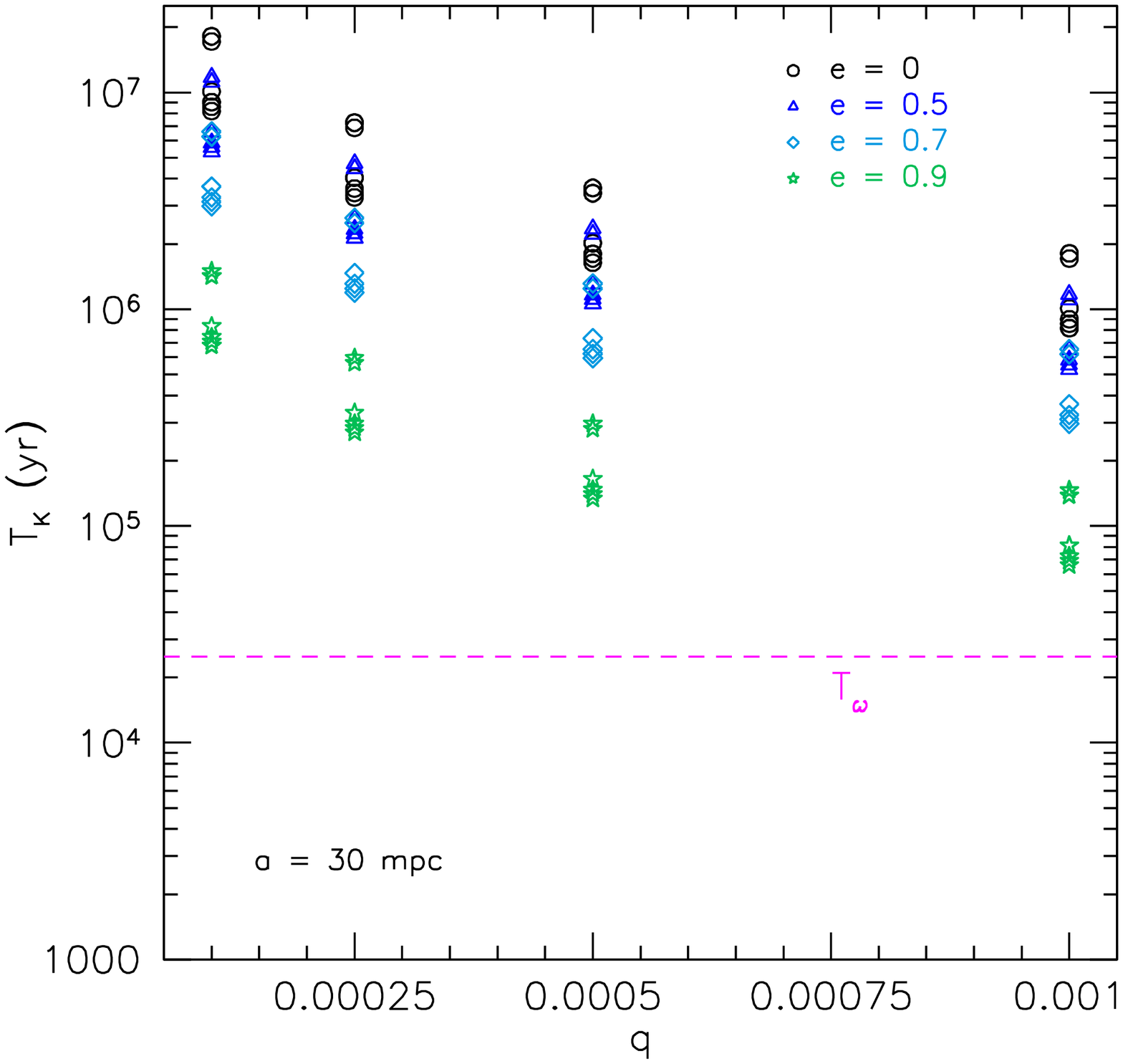}
  \end{center}
  \caption{Timescale for Kozai oscillations in star S2 as a function of
    the black hole binary mass ratio for different values of the
    binary eccentricity. The left panel refers to simulations with
    $a=10\mpc$ while the right panel is for $a=30\mpc$. The horizontal
    lines marks the GR precession timescale for S2.}
  \label{fig:kozai}
\end{figure}
Figure~\ref{fig:kozai} shows $T_K$ as a function of the black hole
binary mass ratio for the assumed values of the eccentricity and for
two allowed separations.  We find that $T_K$ is always longer then
S2's GR precession timescale of $2.5\times10^4\yr$ for $a=30\mpc$. For
the remaining case $a=10\mpc$, $T_K$ is short enough only for
$q\simgreat 2.5\times10^{-4}$ and $e\simgreat 0.7$. This restricts the
applicability of the Kozai mechanism to a small subset of our
simulations, in contrast with the results shown in
Figure~\ref{fig:change} and in the following section.  We conclude that
the Kozai mechanism is not the dominant effect producing changes in
orbital $e$ and $i$ in our simulations.  In those cases where it is
potentially relevant, the eccentricity is predicted to oscillate
between $e_{\rm min} = 0.24 - 0.74$ and $e_{\rm max} = 0.89 - 0.99$
over one Kozai period, depending on the values of $\alpha$ and
$\varpi$.  This implies variations of the order of $2 \times 10^{-5} -
6\times 10^{-3}$ over the integration time of 50 years (for
$a=10\mpc$).

The apoapsis shift due to perturbations from the IBH is very sensitive
to the binary parameters. In the case of $q \simgreat 5\times10^{-4}$
and $a \simless 3\mpc$, the shift over one revolution due to the IBH
becomes larger than the relativistic shift. This suggests a variation
in the orbital elements which is potentially observable with current
instrumentation.

However, the observability of variations in the orbit of S2 depends on
several factors. In the following section we thoroughly examine all
such factors and use orbital fitting to determine whether an IBH is
detectable via on-going monitoring of the S-cluster.  Theoretically,
it would be possible to use other S-stars to investigate the effects
of a hypothetical IBH.  Given that the shifts in the apparent
location of periapsis and apoapsis depend only on the eccentricity and
not on the semi-major axis of the stellar orbit, it would seem
appropriate to consider all stars with $e \simgreat 0.8$ for such an
analysis.  From an observational point of view, however, S2 is the
only star in the sample which is bright enough and not affected by
confusion to allow for meaningful tests of the gravitational
potential. We therefore limit our study to star S2.

\section{Orbital fitting}
In this section we extract observational-like data from the simulated
orbital traces of S2. We assume that eight or nine astrometric epochs
can be obtained each year over the course of 50 years. The eight or
nine yearly epochs are not evenly distributed but are spread over
seven months only, thus taking into account the fact that the Galactic
Centre is accessible with NIR observations only for part of the
year. At the chosen epochs, the original, simulated positions are
perturbed by an astrometric error, assumed to be distributed in a
Gaussian fashion. We use a value of $300\,\mu$as per coordinate, which
is a conservative assumption for S2 \citep{Fritz2010}. The statistical
uncertainty of the measured S2 positions is smaller than the assumed
value. However, unrecognised confusion events with fainter stars are
an additional, important error source, such that we consider our value
realistic.  Furthermore, we assume that the radial velocity of S2 is
determined at two epochs per year. Here, we adopt a Gaussian error of
$15\kms$. This value is typically reached with current NIR
medium-resolution spectrographs.

In this way we obtain 960 simulated data sets that are reasonably
close to what one would obtain by simply continuing the monitoring of
orbits in the Galactic Centre with existing instruments. The fact that
within the next 50 years new facilities with improved angular
resolution will become available (such as the NIR interferometers
ASTRA \citep{astra2008} and GRAVITY \citep{eisen08}, or the extremely
large telescopes TMT and E-ELT) means that an IBH probably will be
detectable more easily than what we derive here. There is a second
reason why we consider our procedure conservative. Observers may adopt
a sampling strategy for the orbits other than a simple constant rate,
as we assume here.  Given the fact that the most constraining part of
an orbit is the periapse passage, and the event is predictable, an
intensification of the observations around the periapse passage
provides an improved sampling pattern.

We fit each of the 960 data sets with the same code as used in
\citet{Gillessen2009}. From the fits, we determine the full set of 13
parameters describing the orbits and the potential: The six orbital
elements $(a,\,e,\,i,\,\Omega,\,\varpi,\,t_P)$ and seven parameters
describing the gravitational potential of the MBH: Mass, distance,
on-sky position (2 parameters) and velocity (3 parameters) of the MBH.
While in the simulations these quantities are known, this is not the
case when the data sets are considered as mock observations. Then
these quantities have to be treated as free fit parameters, since they
need to be determined from the same orbital data from which the
presence of an IBH shall be judged. Currently, nearly all constraints
on these parameters actually come from the S2-orbit which we consider
here.  In future, some of the parameters describing the MBH might be
determined independently from the orbit of other S-stars. We neglect
this here and keep all seven parameters completely free. This is
conservative, because additional constraints would make the presence
of an IBH more easily detectable.

Using a purely Keplerian point-mass model is inadequate for all our
data sets. The relativistic precession is too large during the 50
years of evolution and therefore we include the first order PN
correction to the equations of motion when fitting the orbits
\citep{Gillessen2009}.  For example, all twelve data sets with $(q=1.0
\times 10^{-3},\, a = 30\,\mathrm{mpc},\, e=0.9)$ are well fit by the
relativistic equations.  In contrast, none of these data sets can be
described by a purely Keplerian point-mass potential.  Still, the
relativistic potential only yields a perfect fit for some of the data
sets. For others, the IBH perturbs the dynamics too strongly, and
would therefore be detectable.  Given that our sampling in radial
velocity is rather sparse, the inclusion of special relativistic
corrections to the radial velocities is not needed.

\begin{figure}
  \begin{center}
    \includegraphics[width=8cm]{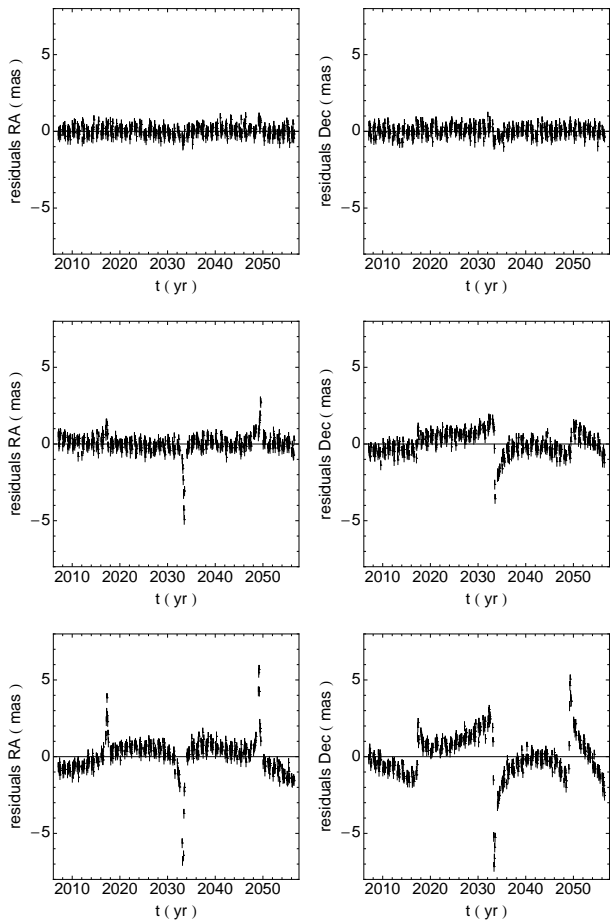}
  \end{center}
  \caption{Examples of fit residuals. Top row: A fit using a simulated
    data set with $(q=1.0 \times
    10^{-4},\,a=3\,\mathrm{mpc},\,e=0)$. Middle row: a fit with
    $(q=2.5 \times 10^{-4},\,a=0.3\,\mathrm{mpc},\,e=0.9)$. Bottom
    row: a fit with $(q=5 \times 10^{-4},\,a=1\,\mathrm{mpc},\,e=0)$.
    The first (top) one is classified as acceptable and has a reduced
    $\chi^2 = 1.15$. The other two are classified as not acceptable
    and have reduced $\chi^2$ values of 4.4 and 22.7 respectively.}
  \label{fig:residuals}
\end{figure}

Note that it is not legitimate to assess the goodness of a fit by
comparing the obtained parameters with those used as input for the
simulations. Instead, we use the reduced $\chi^2$ for each of the 960
fits to decide whether a fit is acceptable or not. We obtain values
between 0.88 and 357. In addition, we examine the residuals of each
fit by eye, dividing them into two categories: A set of fits for which
the residuals do not visually show obvious correlations and a set for
which it is apparent that the chosen gravitational potential model is
not adequate. Figure~\ref{fig:residuals} shows typical examples of the
residuals from three simulations.

Each of the simulated data sets has $418\pm 1$ astrometric data points
and $142\pm 1$ radial velocity measurements. The exact numbers vary due
to the random variations of the sampling pattern. Given that our fits
have 13 free parameters, the number of degrees of freedom is
$n_\mathrm{dof}\approx 2\times 418+142-13=965$.

\begin{figure}
  \begin{center}
    \includegraphics[width=6cm]{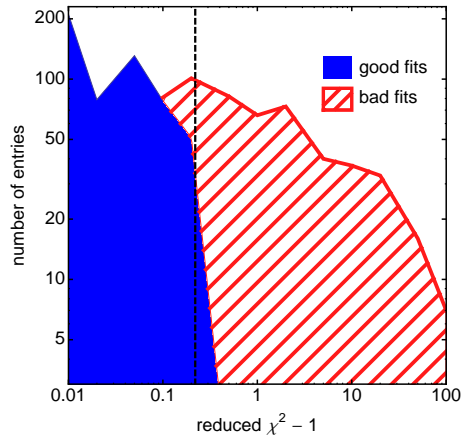}
  \end{center}
  \caption{Distribution of reduced $\chi^2$ for the 960 fits in
    logarithmic bins. Entries are coloured according to our visual
    classification of whether the residuals of any given fit is
    acceptable (blue/solid) or not (red/hatched). The black, dashed
    line marks the optimum cut at 1.22 separating good from bad fits
    by the value of their reduced $\chi^2$.}
  \label{fig:dist}
\end{figure}

Figure~\ref{fig:dist} shows the distribution of reduced $\chi^2$
together with the flag whether a given fit is acceptable or
not. Clearly, the reduced $\chi^2$ can be used as discriminator. The
largest reduced $\chi^2$ corresponding to a fit classified as
'acceptable' is 1.36, the smallest reduced $\chi^2$ corresponding to a
fit classified as 'not acceptable' is 1.12. The optimum cut is at a
reduced $\chi^2$ of 1.22, yielding as many good fits above as bad fits
below the threshold. 
The total number of fits misclassified by this
cut is 41. The total number of bad fits is 409 and correspondingly 551
fits have a reduced $\chi^2$ below the threshold. Hence, the IBH
would be detectable from the data in $\approx$43\% of the cases.  We
find a dependence of the fraction of bad fits on the assumed mass for
the IBH. For $q=10^{-4}$, $2.5\times 10^{-3}$, $5\times 10^{-3}$,
$10^{-3}$ the percentage of detectable IBHs is 15\%, 39\%, 51\%, and
66\%, respectively.

Having a reduced $\chi^2 \geq 1.22$ corresponds formally to a
$4.7\sigma$ detection of the effects of the IBH for
$n_\mathrm{dof}=965$.  In reality, there will be perturbing effects
such as confusion events, recognised or unrecognised. Hence, the
actual $\chi^2$ might be larger than in the simulations and the simple
Gaussian statistics will not apply. We are nevertheless confident that
by observing the astrometric (and photometric) residuals as a function
of time the effects of an IBH can be disentangled from confusion
events. The patterns of the residuals contain more information than a
single $\chi^2$ value.

\begin{figure}
  \begin{center}
    \includegraphics[width=8cm]{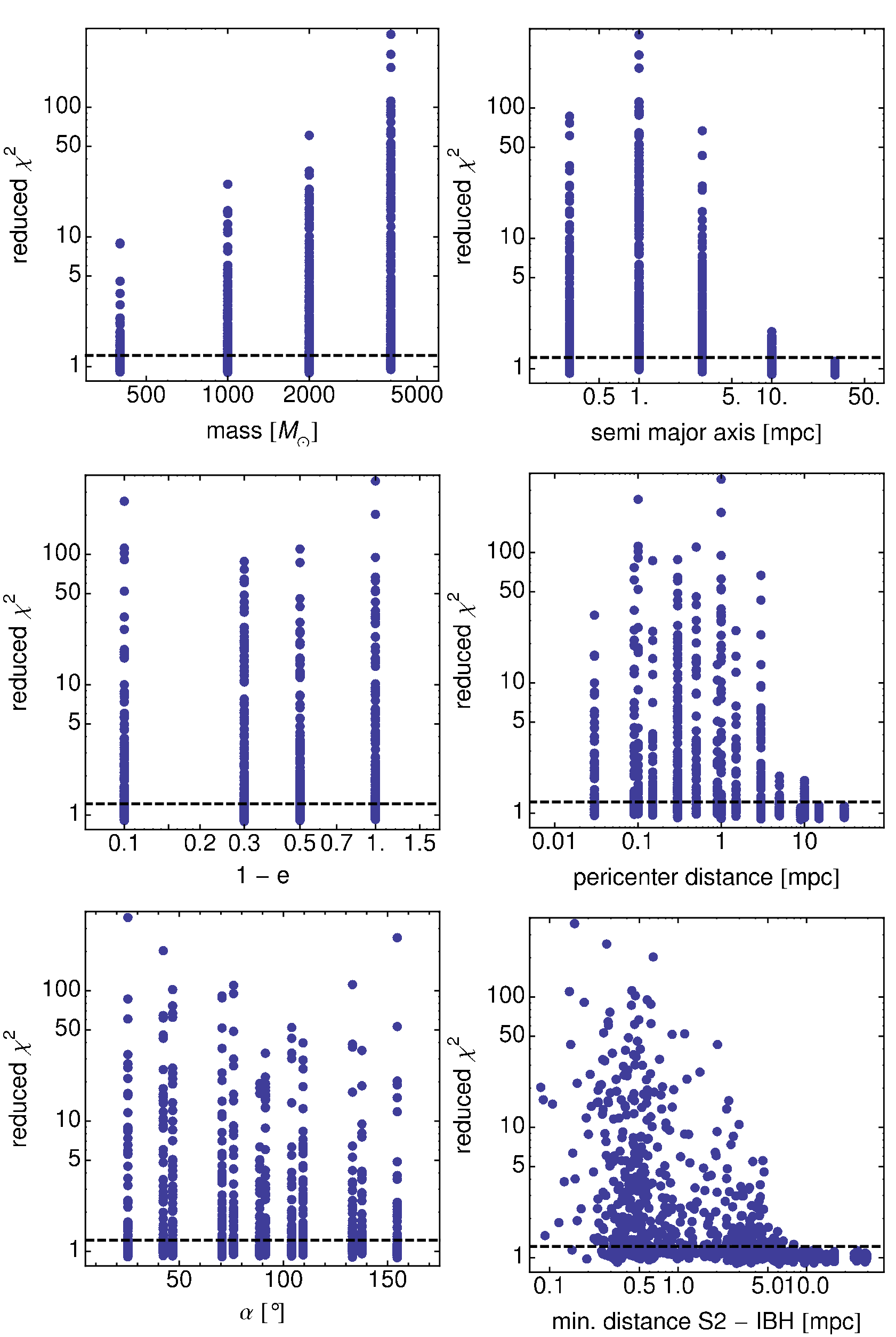}
  \end{center}
  \caption{Distributions of reduced $\chi^2$ for the 960 fits as a
    function of the black hole binary initial parameters: mass,
    semi-major axis, eccentricity, periapse distance, orbital
    orientation and minimum distance to S2.}
  \label{fig:dist2}
\end{figure}
In Figure~\ref{fig:dist2} we examine the dependence of the reduced
$\chi^2$ on the black hole binary parameters. The left panel in the
first row shows that the fits on average get worse when the mass of
the assumed IBH is increased, as expected. The right panel in the
first row shows that a value for the semi major axis of the IBH
around $\approx 1\mpc$ leads on average to the worst fits.
\begin{figure}
  \begin{center}
    \includegraphics[width=6cm]{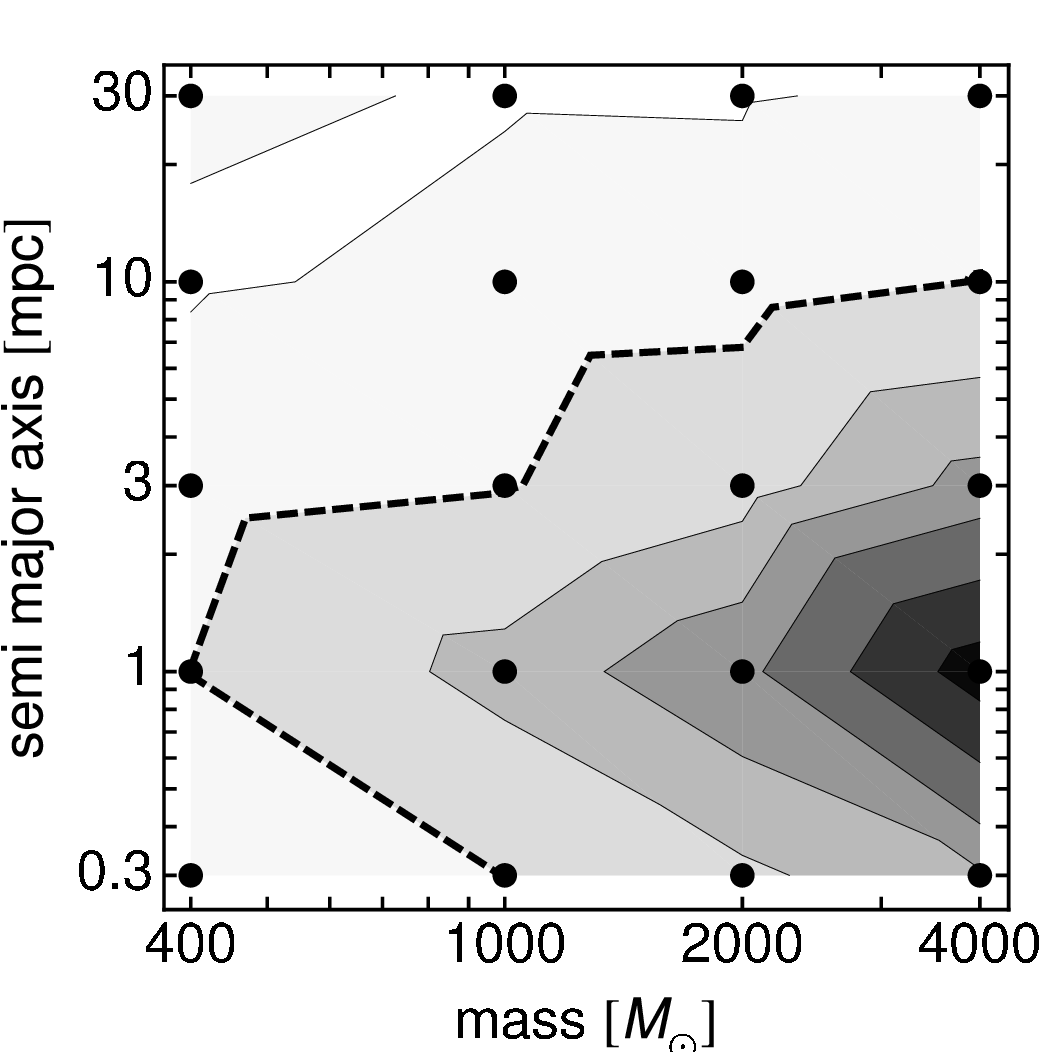}
  \end{center}
  \caption{Reduced $\chi^2$ for the 960 fits as a function of mass and
    semi major axis of the IBH. The plot shows the median at each grid
    point of the $\chi^2$ values. The thick dashed line marks our
    threshold of $\chi^2 = 1.22$. Fits right of the line are not
    acceptable and thus the IBH would be discoverable. }
  \label{fig:dist3}
\end{figure}
 These two parameters correlate actually best with the reduced
 $\chi^2$, and in Figure~\ref{fig:dist3} we show the reduced $\chi^2$
 in the $M_\mathrm{IBH}$-$a$-plane. The unacceptable fits occupy a
 well-defined region in this plot.

The goodness of the fits, on the other hand, is fairly independent of
the eccentricity of the IBH orbit (second row of
Figure~\ref{fig:dist2}, left panel).  The right panel in the second
row shows that the periapsis distance $p=a\,(1-e)$ of the IBH is a
less good predictor for the fit than the semi major axis. The reduced
$\chi^2$ does not correlate with the finite set of orbital
orientations probed, as shown in the third row, left panel of
Figure~\ref{fig:dist2}. Here, $\alpha$ represents the angle between
the angular momentum vectors of S2 and the IBH at the start of the
simulations. This incidentally argues against the possibility that the
observed deviations are due to Kozai oscillations, since the mechanism
requires large relative inclinations to operate.  However, fits with
the same sets of parameters $(q,\,e,\,a)$ but different $\alpha$ can
have different values of $\chi^2$.  This means that the knowledge of
$(q,\,e,\,a)$ is insufficient to predict the reduced $\chi^2$ but
information on the sky position is necessary. Finally, the right panel
in the third row of Figure~\ref{fig:dist2} shows that the minimum 3D
distance between S2 and the IBH also correlates with the reduced
$\chi^2$. In general it holds that the smaller the minimum distance,
the worse the corresponding fit. Clearly, this parameter is not
independent of the semi-major axis.

The initial parameters adopted for the black hole binary are not sampled
homogeneously. This is obvious for the first three panels in
Figure~\ref{fig:dist2} showing the reduced $\chi^2$ as a function of
the binary parameters $(q,\,e,\,a)$. But it also holds (and is less
obvious) for the plot investigating the minimum distance between S2
and the IBH.

\begin{figure}
  \begin{center}
    \includegraphics[width=4cm]{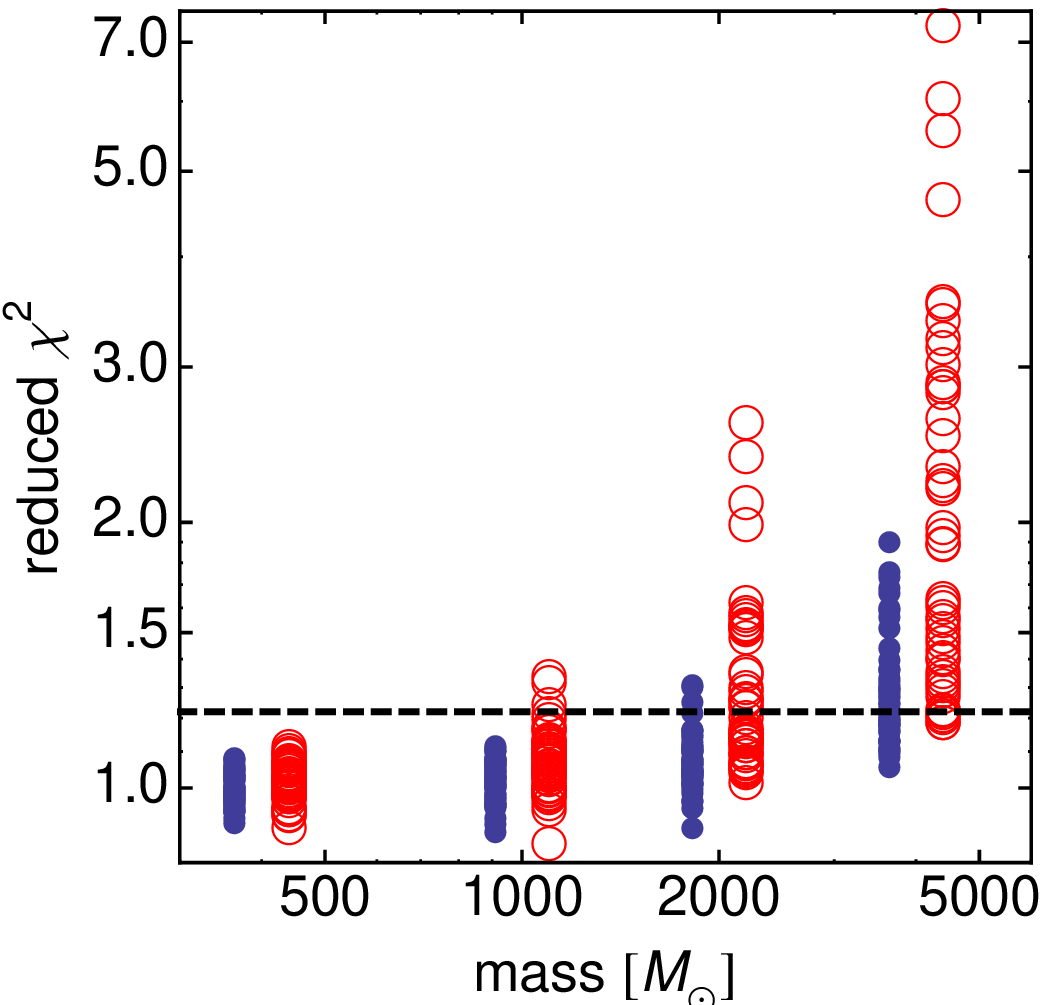}
    \includegraphics[width=4cm]{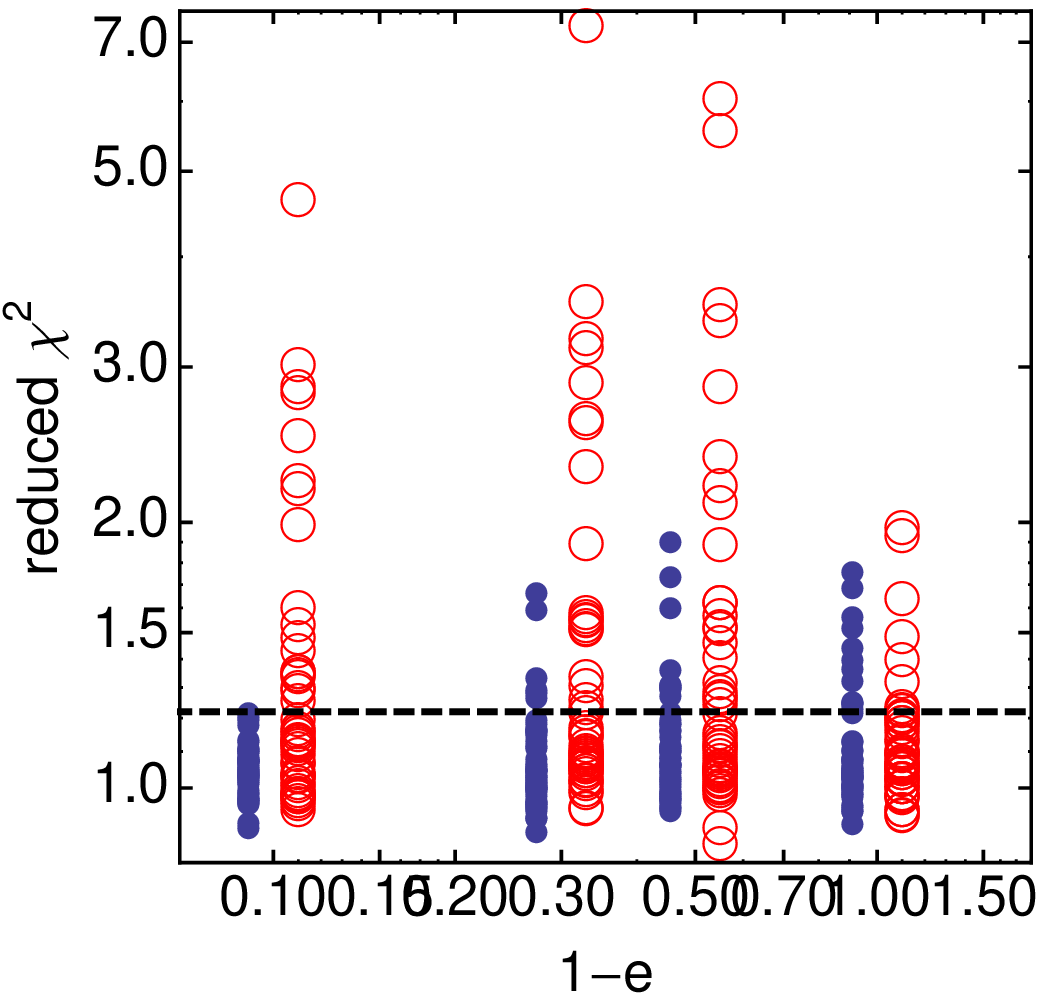}
  \end{center}
  \caption{Reduced $\chi^2$ as a function of black hole mass (left)
    and eccentricity (right) for two set of runs with the IBH starting
    at periapsis (dots) and apoapsis (circles).}
  \label{fig:cfr}
\end{figure}
A comparison of the goodness of fit for the runs starting with 
the IBH at periapsis and apoapsis is shown in Figure~\ref{fig:cfr}.
In both cases, the semi-major axis of the black hole orbit is 
$a=10\mpc$. We find a modest worsening of the $\chi^2$ in the case
of an IBH initially at the apoapsis of its orbit. This can be
attributed to the fact that S2's apoapsis, where the star spends most
of its time, is about $10\mpc$.

\begin{figure}
  \begin{center}
    \includegraphics[width=6cm]{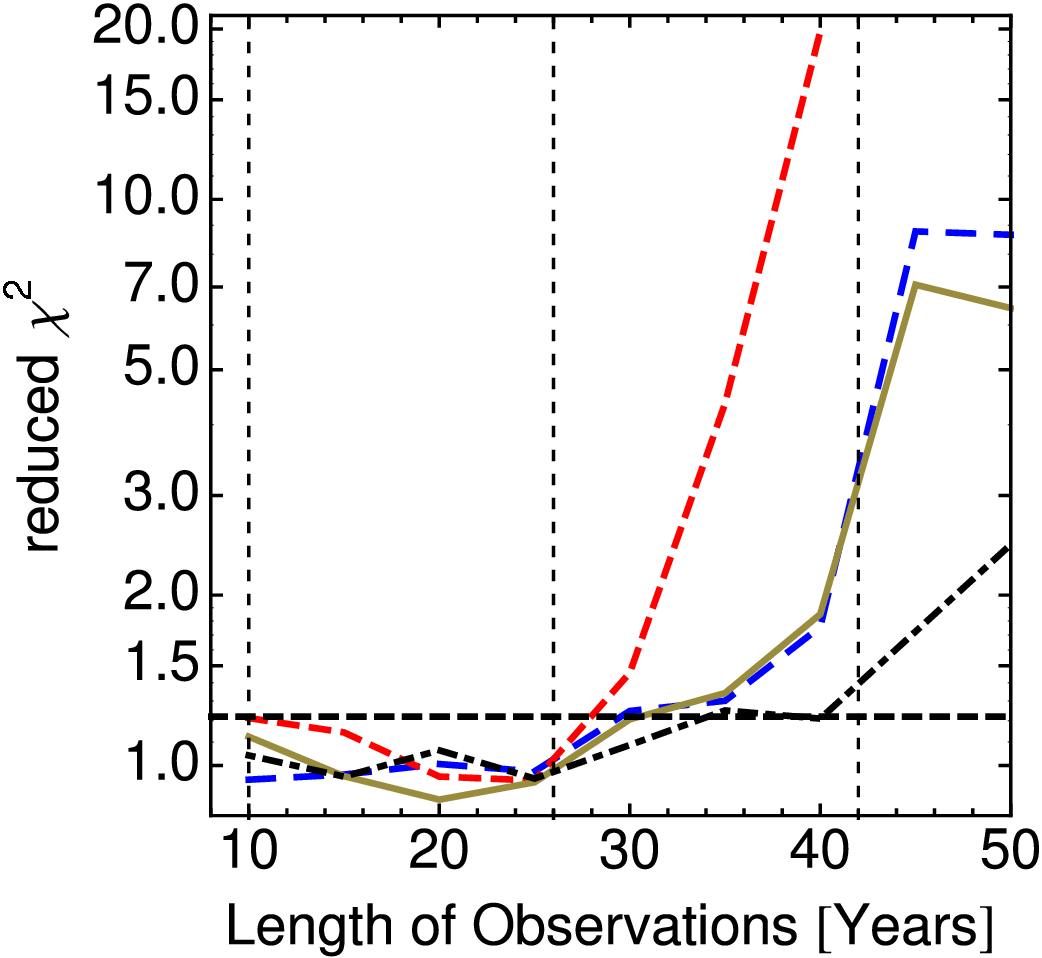}
  \end{center}
  \caption{Reduced $\chi^2$ as a function of the length of the
    observations for four cases: blue (long-dashed): $q=5 \times
    10^{-4}$, $e=0.9$, $a=1\mpc$; red (dashed): $q=10^{-3}$, $e=0.9$,
    $a=1\mpc$; green (solid): $q=10^{-3}$, $e=0$, $a=0.3\mpc$; black
    (dot-dashed): $q=10^{-3}$, $e=0.7$, $a=3\mpc$. The vertical lines
    indicate the second and third periapse passage covered by the
    simulations, the first one happens at $t\approx 10\yr$. The
    horizontal, dashed line is our optimum cut at 1.22.}
  \label{fig:time}
\end{figure}

Finally, we also investigated the minimum time required for the IBH
to become detectable. For this purpose, we repeated the orbital fits
for a few cases assuming that the observations span 10,15,
20,25, 30, 35, 40, 45 or 50 years (Figure~\ref{fig:time}). 
Our initial conditions are such that the first periapse passage of S2
happens after 10 years, the second after 26 years and the third after
42 years. Figure~\ref{fig:time} shows that the reduced $\chi^2$ starts
to increase beyond our threshold of 1.22 after the second periapsis
passage for fits that show a large reduced $\chi^2$ after 50 simulated
years (red/dashed, blue/long-dashed and green/solid curves). Only for
fits that after 50 simulated years have a reduced $\chi^2 \simless 3$
is the threshold passed after the third periapse passage (black
curve). The discrete nature of periapse passages also is the reason
why the blue/long-dashed and green/solid curves level out after about
45 years.

This is interesting in comparison with the current status of the
observations. Since by now only one periapse passage of S2 has been
observed in 2002, one would not expect to have detected an IBH from
the actual S2 data so far. This is particularly true since the assumed
level of accuracy was not reached during the first years of the
observations (1992 - 2002), and radial velocity information is only
available after 2002 (with the exception of one point in 2000). Hence,
the first real chance to detect an IBH will be after the next S2
periapse passage, which will happen in 2018.

\section{Other sources of orbital torque}
\label{sec:torque}

Here we consider other perturbations that could induce changes
in S2's orbital angular momentum, potentially complicating the
signal from an IBH.
We find that almost all such alternative perturbations are small
compared to the torque produced by an IBH.

Angles quoted in this section are intrinsic, not astrometric.
As noted above, current  data are able to determine the
orbital angles ($\varpi,\Omega,i$) of S2 with an accuracy of
about one degree.
Changes induced by an IBH per orbit of S2 are 
$10^{-3} \mathrm{deg} \simless \Delta(i,\Omega) \simless 1 \mathrm{deg}$
(Figure~2).

{\em Spin of the MBH.}
Frame-dragging effects from a spinning MBH include an
additional in-plane precession term as well as a precession of the
orbital plane.  Defining $i^\prime$ and $\Omega^\prime$ as
the inclination and nodal angle of S2's orbit 
with respect to the MBH's equatorial plane, 
to lowest PN order, frame dragging induces changes
\begin{subequations}
\begin{eqnarray}
\Delta\varpi &=& -2A_J\cos i^\prime, \\
\Delta\Omega^\prime &=& A_J
\end{eqnarray}
\end{subequations}
per revolution in the angle of periapse and the line of nodes,
respectively, where
\begin{subequations}
\begin{eqnarray}
A_J &=& \frac{4\pi\chi}{c^3} \left[\frac{GM_\mathrm{MBH}}{(1-e^2)a}\right]^{3/2} \\
&\approx& 0.115' \left(1-e^2\right)^{-3/2}\chi \left(\frac{a}{\mathrm{mpc}}\right)^{-3/2}
\end{eqnarray}
\end{subequations}
and $\chi\le 1$ is the dimensionless spin of the MBH \citep[e.g.][]{MAMW2010}.
The orbital inclination $i^\prime$ remains unchanged.
For S2, the spin contribution to advance of the periapse is $\sim 1\%$ of the
Schwarzschild contribution even for $\chi=1$ and so can be ignored.
The nodal advance is $\Delta\Omega^\prime \approx 0.002\,\chi\,\rm deg$, 
too small to be detectable in the next few decades of monitoring,
and smaller than the changes induced by an IBH in almost all of the
runs carried out here (Fig.~2).
Effects of frame dragging are only likely to be important for stars 
much closer to SgrA$^*$ than S2 \citep{MAMW2010}.
(Frame dragging could nevertheless be relevant to the orbit of an IBH.
For the IBH orbit with smallest
$a$ ($0.3$ mpc) and largest $e$ ($0.9$) considered here,
the precession time drops to $\sim 300\chi^{-1}$ yr.)

{\em A stellar bar.} If the gravitational potential due to the
distributed mass is appreciably non-spherical, the resultant torques
could affect all of the orbital elements of S2 aside from $a$.
For instance, the nuclear bar described by \citet{alard01} 
has been modelled as a triaxial spheroid with central density
$\sim 150 \msun \mathrm{pc}^{-3}$ \citep{combes08}.
In the most extreme case, the nuclear star cluster (NSC) 
of the Milky Way could
be stratified on triaxial ellipsoids at all radii.
A homogeneous, non-rotating triaxial bar induces changes in the
inclination and nodal angle of a test star orbiting near the MBH,
with characteristic time scale
\begin{equation}
T_{\Omega,i}\approx \frac{\sqrt{1-e^2}}{T_i-T_j} \frac{1}{P G\rho_b}
\end{equation}
\citep[][equations 11-15]{MV10}.
Here, $\rho_t$ is the density of the triaxial component,
and $(T_x,T_y,T_z)$ are the dimensionless coefficients,
of order unity, that define the shape of the triaxial component
\citep{chandra69}; the torque in the $(i,j)$ principal plane is 
proportional to $T_i-T_j$, etc.
For S2, the angular reorientation over one orbit due to torques
from a triaxial bar would be of order
\begin{equation}
\Delta\phi\approx 0.2\mathrm{mas}\ (T_i-T_j)
\left(\frac{\rho_b}{100 \msun \mathrm{pc}^{-3}}\right),
\end{equation}
undetectable even if $\rho_b\approx 10^5\msun \mathrm{pc}^{-3}$,
the approximate density of the NSC at $1$ pc from SgrA$^*$.

{\em Resonant relaxation:}
Discreteness in the distribution of stars and stellar remnants
is also a potential source of torque.
``Vector resonant relaxation'' produces changes in the
direction of the orbital angular momentum of order 
\begin{equation}
\Delta\phi \approx \frac{\sqrt{N}m}{\mbh} 
\end{equation}
per radial period,
where $m$ is the mass of a background star and $N$ is the number
of such stars within S2's orbit.
Writing $N=M_\star/m$, with $M_\star$ the distributed mass
within S2's orbit, this is
\begin{eqnarray}
\Delta\phi &\approx& \left(\frac{M_\star}{\mbh}
\frac{m}{\mbh}\right)^{1/2}  \\
&\approx& 0.003\ \mathrm{deg} 
\left(\frac{M_\star/\mbh}{10^{-2}}\right)^{1/2}
\left(\frac{\mbh}{4\times 10^6m}\right)^{-1/2}.  
\end{eqnarray}
Since $M_\star\simless 10^{-2}\mbh$ \citep{gillS2},
this is smaller even than the change due to frame-dragging
unless the mean perturber mass $m\gg\msun$.

\section{Discussion and conclusions}
We have shown that an IBH orbiting Sgr A$^*$ at the distance 
of the S-stars can cause observable deviations in the orbit of star S2. 
The perturbations potentially affect all the orbital elements, 
but the key signature would be a change in S2's angular momentum 
(eccentricity, orbital plane) due to the non-spherically-symmetric 
forces from the IBH. 
In particular, we find that an IBH more massive
than $\sim 1000\msun$ at a distance of $\sim 1-5\mpc$ is potentially
discoverable at the next periapse passage of S2.

\begin{figure}
  \begin{center}
    \includegraphics[width=8cm]{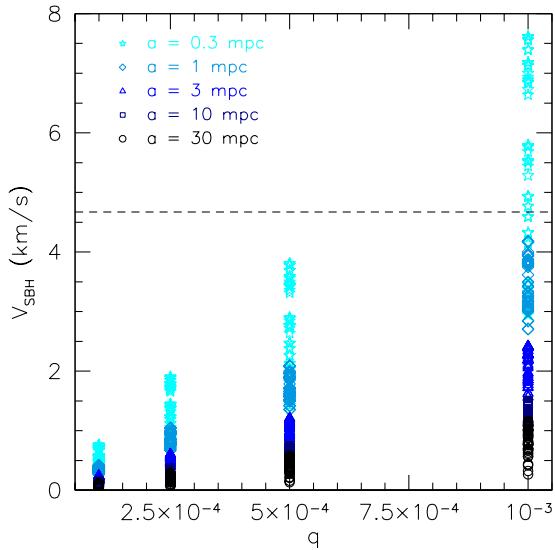}
  \end{center}
  \caption{Root mean square velocity of the MBH as a function of the
    binary mass ratio. The dotted line represents the $3\sigma$ limit
    on the 3D proper motion of SgrA* derived from \citet{r09}.}
  \label{fig:vbh}
\end{figure}

The presence of an IBH companion to the MBH might produce an
observable feature in the radio observations of SgrA*.  In particular,
some combinations of the binary parameters considered here might
produce a reflex motion of the MBH which is larger than the currently
available limits on the proper motion of SgrA*. Such combinations of
parameters could therefore be considered unlikely.  \citet{r09} report
a proper motion of $\left(7.2 \pm 8.5\right)\kms$ in the plane of the
Galaxy and of $\left(-0.4\pm0.9\right)\kms$ in the direction
perpendicular to the plane. Since it appears unlikely that the motion
of the MBH lies primarily in the Galactic plane, we adopt the value of
the perpendicular component of the velocity as our fiducial value. In
Figure~\ref{fig:vbh} we compare the 3D root mean square velocity of
the MBH obtained from the simulations with the observational $3\sigma$
upper limit.
We find that the motion of the MBH induced by the orbiting IBH is
generally smaller than the $3\sigma$ limit derived from radio
observations of SgrA*. Only for $q > 10^{-3}$ and $a=0.3\mpc$ does the
simulated motion exceed the limit. Of course, if the motion induced by
the IBH were to lie primarily in the Galactic plane, it would be very
hard to detect via radio observations, even for large values of $q$.

\section*{Acknowledgments}
DM acknowledges support from the National Science Foundation under
grants no. AST 08-07910, 08-21141 and by the National Aeronautics and
Space Administration under grant no. NNX-07AH15G.


\label{lastpage}

\end{document}